\documentclass[twocolumn,amssymb,nobibnotes,superscriptaddress,aps,pra,tightenlines]{revtex4-1}
\usepackage{color,graphicx,amsfonts,amssymb,amsmath,footmisc,epsfig,fancyhdr,subfigure,braket}
\usepackage[colorlinks=true, linkcolor=red, citecolor=blue, urlcolor=blue]{hyperref}
\begin{document}

\title{Non-Abelian Gauge Enhances Self-Healing for Non-Hermitian Su-Schrieffer-Heeger Chain}
\author{Yazhuang Miao}
\affiliation{ School of Science, Qingdao University of Technology, Qingdao, Shandong, China }
\author{Yiming Zhao}
\affiliation{ School of Science, Qingdao University of Technology, Qingdao, Shandong, China }
\author{Yong Wang}
\affiliation{ School of Science, Qingdao University of Technology, Qingdao, Shandong, China }
\author{Jie Qiao}
\affiliation{ School of Science, Qingdao University of Technology, Qingdao, Shandong, China }
\author{Xiaolong Zhao$^{*}$}
\affiliation{ School of Science, Qingdao University of Technology, Qingdao, Shandong, China }
\author{Xuexi Yi$^{**}$}
\affiliation{Center for Quantum Sciences and School of Physics, Northeast Normal University, Changchun, Jilin, China}
\date{\today}

\begin{abstract}

We investigate a non-Hermitian extension of the Su-Schrieffer-Heeger model that incorporates spin-dependent
SU(2) gauge fields, represented by non-Abelian couplings between lattice sites, as well as independent
nonreciprocal hopping amplitudes. This framework gives rise to a rich phase structure characterized by
complex-energy braiding and tunable non-Hermitian skin effects. By employing the generalized Brillouin zone
approach, we analyze the bulk-boundary correspondence and identify topological transitions protected by chiral
symmetry. Notably, we demonstrate that non-Abelian gauge fields significantly enhance the dynamical resilience of the system,
enabling robust self-healing under a moving scattering potential. These results clarify the role of SU(2) gauge fields
in stabilizing non-Hermitian topological phases and indicate that the proposed model can be realized with
currently available photonic, atomic, and superconducting experimental platforms.

\end{abstract}

\maketitle
\thispagestyle{fancy}
\lhead{}
\cfoot{}
\rfoot{}

\section{Introduction}
Topological phases of matter can be distinguished by global invariants, giving rise to phenomena with significant
implications for quantum computation, resilient quantum transport, and spin-based technologies~\cite{Hasan2010, Qi2011, Chiu2016}.
In Hermitian systems, the topological invariants predict the existence of boundary-localized states which are
protected against local perturbations. Recent developments in experimental platforms, such as ultracold atoms~\cite{Goldman2014},
photonic systems~\cite{Pereira20246}, and magnonic lattices~\cite{Zhang2017CavityMagnonEP}, have extended the research of
topological phases into non-Hermitian settings, where gain, loss, and nonreciprocal couplings emerge as intrinsic features~\cite{ElGanainy2018, Ashida2017, Shen2018}.
Non-Hermitian Hamiltonians host novel spectral and dynamical phenomena absent in Hermitian counterparts, including
exceptional points~\cite{Lee2016, Leykam2017, Ding2018, Miri2019}, spectral winding, and the non-Hermitian skin effect
(NHSE)~\cite{Yao2018,Gong2018,Shen2025NHSEonQPU,1Manna20236} wherein extensive numbers of eigenstates
accumulate at the boundaries of a system. This breakdown of conventional bulk-boundary correspondence has prompted the
development of generalized theoretical tools, notably the generalized Brillouin zone (GBZ), which extends the Bloch
framework into the complex plane by analyzing the characteristic polynomial of the non-Hermitian Bloch Hamiltonian~\cite{Yokomizo2019,1Mandal2024135}.
While these theoretical advances broaden the topological classification of non-Hermitian systems, they also highlight
new sensitivities: the same mechanisms that enable directional amplification also render the spectrum fragile under
scattering, loss, or fabrication-induced disorder~\cite{1Mandal2025111,1Gangaraj20202}. Interestingly, recent studies have observed that
certain non-Hermitian systems can exhibit a self-healing response, in which localized modes partially recover after perturbations, suggesting
a possible route to mitigate the fragility of non-Hermitian edge modes~\cite{LonghiSelfHealinPRL}. However, this recovery tends to be highly
sensitive to system parameters, limiting its practical robustness.
In this work, we show that by introducing spin-dependent SU(2) gauge fields,
the self-healing dynamics can be significantly enhanced and stabilized, resulting in tunable and resilient behavior even under strong space-time-dependent
scattering potential.

The Su-Schrieffer-Heeger (SSH) chain has served as a cornerstone for studying one-dimensional
topological phases. Its non-Hermitian extensions have revealed a host of novel spectral features~\cite{Lee2016,Yao2018,LonghiSelfHealinPRL,1Sarkar202426}.
Across a range of physical platforms, including optics~\cite{Zeuner2015} and acoustics~\cite{Hu2023},
non-Hermitian generalizations of the SSH model have been experimentally explored, further emphasizing the role of
non-Hermiticity in shaping topological behavior~\cite{ElGanainy2018}.

Simultaneously, synthetic gauge fields, both Abelian and non-Abelian, have emerged as powerful tools for engineering
spin-orbit coupling in a wide variety of systems, including cold atomic gases~\cite{Goldman2016, Aidelsburger2018,Goldman2014,Cooper2019} and
photonic lattices~\cite{Aidelsburger2018,Berry1984,Lu2014}. In particular, non-Abelian SU(2) gauge structure can imprint spin-dependent phases
on hopping processes, effectively coupling internal spin degrees of freedom to spatial motion in a manner that can drastically alter the topological
character of the systems~\cite{Yang2020,Q.Liang2022,Arwas2022}.

Recent experimental advances have demonstrated the realization of asymmetric and nonreciprocal couplings across a wide range of platforms, including
optical ring-resonator arrays~\cite{Liu2022LaserArrays}, acoustic systems exhibiting transient NHSE~\cite{Gu2022TransientNHSE},
cavity magnon-polariton structures~\cite{Zhang2017CavityMagnonEP}, and superconducting circuits supporting directional
supercurrents~\cite{He2025AsymSupercurrent, Davydova2024NonreciprocalSC}. In parallel, spin-dependent SU(2) hopping has been engineered via Raman-induced
spin-orbit coupling in ultracold atomic gases~\cite{Rey2016SyntheticGauge, Guo2025DensityGauge}, as well as in metasurfaces and photonic-chip
waveguides~\cite{Zhan2022MetasurfaceSpin,Whittaker202115193}. Techniques for introducing controlled dissipation, such as patterned loss in photonic
lattices~\cite{Pereira20246}, have also been developed. Collectively, these efforts highlight the increasing experimental control over non-Hermitian
and gauge-related phenomena, setting the stage for deeper theoretical exploration of their topological and dynamical consequences.

In this work, we investigate a non-Hermitian generalization of the SSH model in which spin-$1/2$ degrees of freedom are subject to synthetic non-Abelian
SU(2) gauge couplings that introduce spin-dependent hopping amplitudes. The resulting interplay between non-Hermiticity and gauge-induced spin-orbit
interaction leads to a range of distinctive topological features, including complex-energy braiding, nontrivial boundary localization, and spectral transitions.
By employing the GBZ formalism, we establish a consistent bulk-boundary correspondence and define a topological invariant that captures the transitions between
phases in the presence of chiral symmetry. A central result of our analysis is that non-Abelian gauge fields can strongly enhance the dynamical resilience of
the system: certain eigenstates exhibit a robust self-healing response following interaction with time-dependent scattering potentials. Our numerical simulations
show that SU(2) gauge parameters can be precisely tuned to restore the spatial structure of these states even in the presence of significant external perturbations. Throughout, we initialize in the OBC eigenmode with the largest $\mathrm{Im}\,(E)$, since the largest-gain component governs the long-time response; such modes and their self-healing dynamics are routinely accessible in photonic waveguide arrays~\cite{Bai2025,Ma2025}.

This paper is organized as follows. In Sec.~\ref{sec2}, we introduce the non-Hermitian SU(2) SSH model and detail its construction. Section~\ref{sec3} presents
the spectral properties and NHSE phase structure, obtained via exact diagonalization and GBZ analysis. In Sec.~\ref{sec4}, we explore how non-Abelian gauge
fields enhance self-healing dynamics under scattering. Finally, we conclude in Sec.~\ref{sec5} with a discussion of implications and future directions.

\section{Model}
\label{sec2}
\begin{figure}[htbp]
    \centering
    \includegraphics[width=\linewidth]{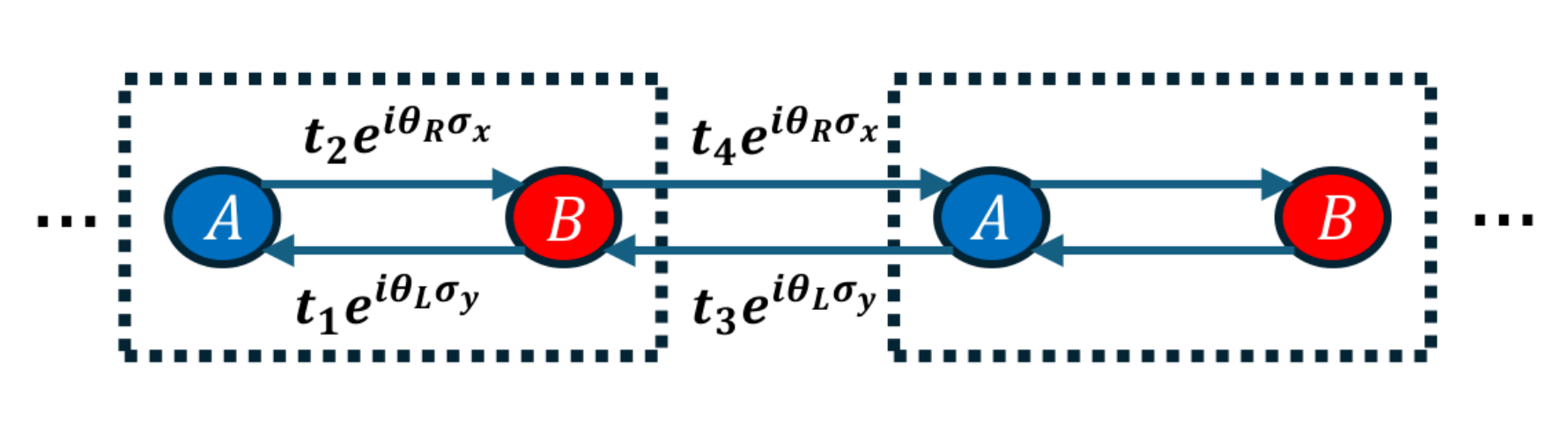}
    \caption{
    Schematic of the non-Hermitian non-Abelian SU(2) SSH model. Each unit cell (dotted box) contains two sublattice sites, labeled $A$ and $B$, and each site hosts a spin-$1/2$ degree of freedom. The hopping amplitudes $t_1, t_2, t_3, t_4$ connect sublattices within
    (and across) unit cells, while the SU(2) rotation matrices $U_L$ and $U_R$ endow these processes with spin-dependent phases.}
    \label{figmodel}
\end{figure}

We consider a non-Hermitian extension of the SSH model that incorporates non-Abelian gauge structures via
SU(2) rotations~\cite{Chen2019,Alexandre2019,Pang2024}. The OBC Hamiltonian reads
\begin{align}
H &= \sum_{n=1}^{N} \Bigl[ t_1\,a_n^\dagger U_L\,b_n + t_2\,b_n^\dagger U_R\,a_n \Bigr] \nonumber\\[1mm]
&\quad + \sum_{n=1}^{N-1} \Bigl[ t_3\,b_n^\dagger U_L\,a_{n+1} + t_4\,a_{n+1}^\dagger U_R\,b_n \Bigr],
\label{eqmodel}
\end{align}
with the annihilation spinor operators on lattice $n$
\begin{align}
a_n = \begin{pmatrix} a_{n\uparrow} \\[2mm] a_{n\downarrow} \end{pmatrix}, \quad
b_n = \begin{pmatrix} b_{n\uparrow} \\[2mm] b_{n\downarrow} \end{pmatrix},
\end{align}
act on sublattice sites $A$ and $B$, respectively. Here, $t_1,t_2$ denote intra-cell hopping amplitudes and $t_3,t_4$
denote inter-cell hoppings. Non-Hermiticity results simultaneously from asymmetric hopping terms as well as the spin-dependent SU(2)
rotations, namely
\begin{align}
U_{s} = \exp\Bigl(i\,\theta_{s}\,\sigma_{s}\Bigr) = \cos\theta_{s}\,\sigma_0 + i\,\sin\theta_{s}\,\sigma_{s},\quad s\in\{L,R\}.
\end{align}
In this work, we identify $\sigma_L \equiv \sigma_y\quad \text{and} \quad \sigma_R \equiv \sigma_x,$ with $\theta_L,\theta_R\in\mathbb{R}$
control the orientation and strength of the gauge-induced spin-orbit coupling. Such a model described by Eq.~\eqref{eqmodel}
is schematically shown in FIG.~\ref{figmodel}.

\section{Multi-type spectra and NHSE}
\label{sec3}
\subsection{Spectral phase diagram}
The Bloch Hamiltonian after Fourier transformation of the Hamiltonian ~\eqref{eqmodel} under PBC reads
\begin{align}
H(k) = \begin{pmatrix}
0 & t_1 U_L + t_4 e^{-i k}U_R \\
t_2 U_R + t_3 e^{i k}U_L & 0
\label{Hk}
\end{pmatrix}.
\end{align}

\begin{figure}[htbp]
\subfigure{\includegraphics[width=\linewidth]{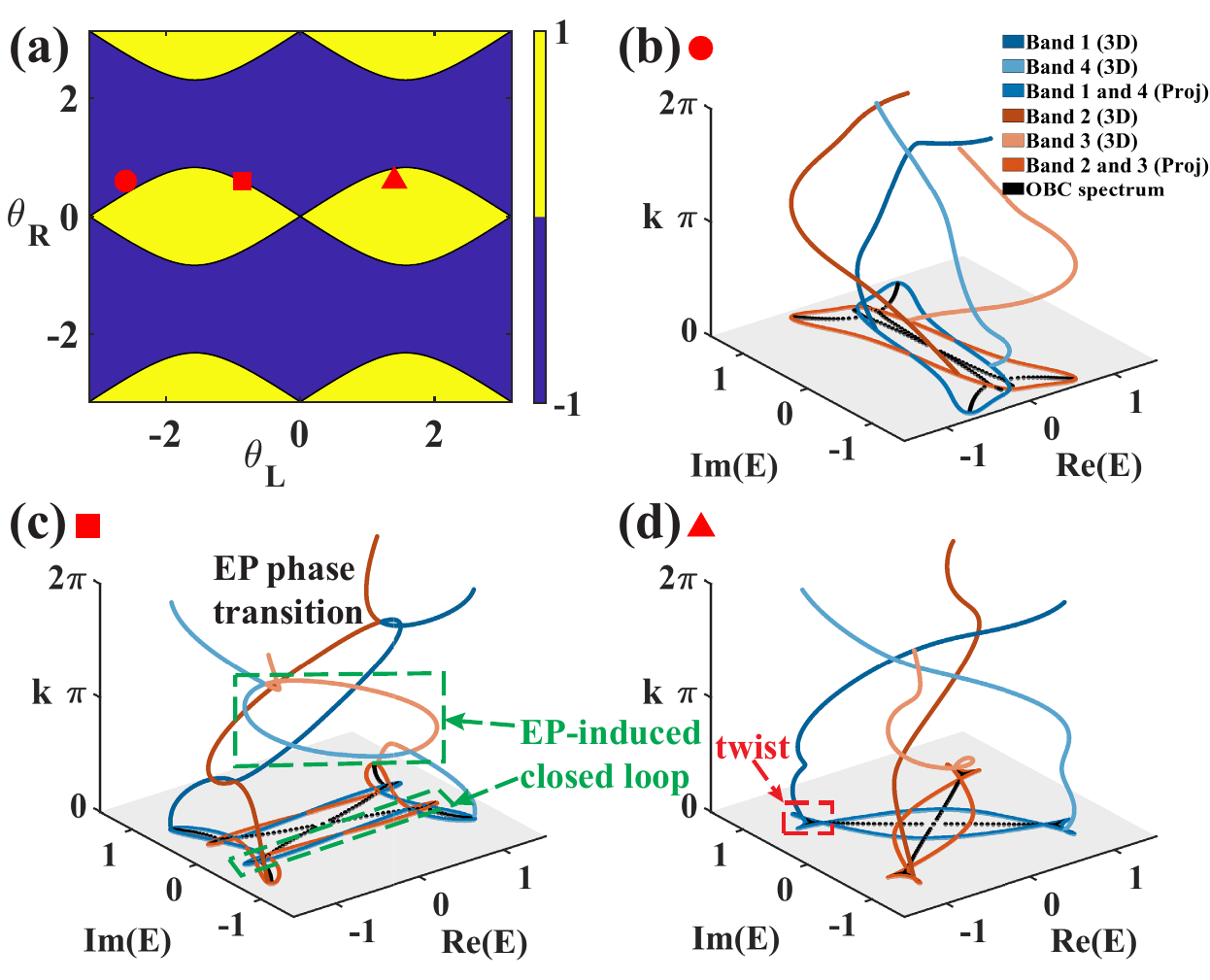}}
\caption{
Phase diagrams, braiding and complex-energy spectrum. (a) Phase diagram versus $\theta_{L}$ and $\theta_{R}$ at $(t_1,t_2,t_3,t_4)=(0.60,1.00,0.80,0.89)$. Points $(\theta_L,\theta_R)=(-2.6,0.6)$, $(-0.874,0.6)$, and $(1.4,0.6)$ correspond to (b), (c), and (d). (b)--(d) Braiding and complex-energy spectra under periodic (color loops) and open boundary conditions (black arcs). In (c), the closed loops in $(\mathrm{Re}\,E,\mathrm{Im}\,E,k)$, as well as their projection onto the complex-energy plane are actually formed by two different bands (EP-mediated). $N=49$ hereafter for calculations about real space Hamiltonians. The legend of (c) and (d) is identical to that of (b).
}
\label{figphaseEnergy}
\end{figure}

The eigenenergy of ${H}(k)$ is given by
\begin{equation}
E(k) = \pm \sqrt{\frac{1}{2}\left[\mathrm{tr}\,M(k) \pm \Delta(k)\right]}.
\end{equation}
where $\quad M(k) \equiv h_k^{(+)}\,h_k^{(-)}$ with
$h_k^{(+)} = t_1 U_L + t_4 e^{-i k}U_R$, and
$h_k^{(-)} = t_2 U_R + t_3 e^{i k}U_L$,
and $\Delta(k)=\sqrt{\left[\mathrm{tr}\,M(k)\right]^2-4\,\det M(k)}$. A second-order exceptional point (EP) arises when two eigenvalues and their eigenvectors coalesce, that is
\begin{equation}
\Delta(k)=0\quad\Longleftrightarrow\quad\bigl[\operatorname{tr}M(k)\bigr]^{2}=4\det M(k),
\label{eq:EPcondition}
\end{equation}
which we adopt as the analytic EP criterion used throughout the discussion of FIG.~\ref{figphaseEnergy} (c).
The chiral (or sublattice) symmetry is encoded by the operator
\begin{equation}
C = \begin{pmatrix} \sigma_0 & 0 \\[2mm] 0 & -\sigma_0 \end{pmatrix},
\end{equation}
as
\begin{equation}
  C\,H(k)\,C^{-1} = -H(k).\label{Eq5C}
\end{equation}
This symmetry guarantees the Hamiltonian being the off-diagonal form of Eq.~\eqref{Hk}.

To characterize the topological phases of this non-Hermitian chain,
we use the braiding degree, a Gauss-linking invariant that counts
the winding of two eigenenergy trajectories in the
$(\text{Re}(E),\text{Im}(E))$ plane as the momentum $k$ evolves~\cite{Wang2021}.
Because bands 1 and 4 (or 2 and 3) permute after $k\mapsto k+2\pi$,
each analytic band is $4\pi$-periodic. Hence the invariant must be
evaluated over $k\in[0,4\pi]$, where two closed trajectories form a Hopf link~\cite{Fukui1998}.
In practice we compute it from two non-degenerate bands whose eigenvalue
trajectories define distinct closed loops in the complex-energy plane.
Such braiding degree $ w = \pm1 $ as shown in FIG.~\ref{figphaseEnergy} (a) can be defined as
\begin{equation}
w = \frac{1}{4\pi}
\oint_{C_1}
\oint_{C_2}
\frac{(\mathbf{r}_1 - \mathbf{r}_2)\cdot\bigl(d\mathbf{r}_1 \times d\mathbf{r}_2\bigr)}{\|\mathbf{r}_1 - \mathbf{r}_2\|^3}\,,
\label{eq:Gauss}
\end{equation}
where $C_{1,2}$ are the trajectories of the chosen bands over $k\in[0,4\pi]$.
The position vectors are
$\mathbf{r}_{1,2}(k) = (\text{Re}[\lambda_{1,2}(k)], \text{Im}[\lambda_{1,2}(k)], k)^T$,
with $\lambda_{1,2}(k)$ the corresponding eigenvalues. Importantly, the closed trajectories enclosed by the green dashed box
visible over $k \in [0,2\pi]$ in FIG.~\ref{figphaseEnergy} (c) are EP-mediated composite loops that close only after
band permutation across the exceptional point, i.e., by concatenating
segments from two distinct non-degenerate bands.
These loops must not be used to evaluate the Gauss-linking invariant.
In contrast, the Hopf link we classify is formed by two separate closed loops in the gray energy plane of
FIG.~\ref{figphaseEnergy} (b)--(c), one shown in red and the other in blue.

As illustrated in FIG.~\ref{figphaseEnergy}, the phase diagram according to
Eq.~(\ref{eq:Gauss}) is shown in panel (a) versus $\theta_{L}$ and $\theta_{R}$.
The solid circle ($\theta_L=-2.6, \theta_R=0.6$), square ($\theta_L=-0.874, \theta_R=0.6$),
and triangle ($\theta_L=1.4, \theta_R=0.6$) are chosen as three representative points for three
phases. Each shows qualitatively different energy braiding and complex-energy spectra as illustrated
in FIG.~\ref{figphaseEnergy} (b)--(d), respectively. In detail, under PBC, the complex-energy spectrum forms closed loops (colored) in the bottom panel, whereas under OBC it collapses into distinct open arcs shown as black curves in the bottom panel. In the
non-Hermitian case, the distinct differences of the energy spectra under open boundary conditions
and periodic boundary conditions significantly deviate from the Hermitian scenario.

Particularly notable is the triangle (FIG.~\ref{figphaseEnergy} (d)) point, where twisting in the
spectral loops appear. This twisting corresponds to two-sided NHSE regimes as shown in FIG.~\ref{figgbzNhse} (f),
namely, not all eigenstates localize at only one end of the chain. In contrast, at the circle
(FIG.~\ref{figphaseEnergy} (b)) and the square (FIG.~\ref{figphaseEnergy} (c)) points, there is no twisting,
reflecting predominantly one-sided localization as shown in FIG.~\ref{figgbzNhse} (b) and (d). Hence,
the degree of twisting in the complex-energy loops directly correlates with the localization behavior
of eigenstates at the boundaries, which can be controlled by adjusting the non-Abelian gauge coupling
parameters $\theta_L$ and $\theta_R$ in this work.
The deeper physical mechanism connecting complex-energy band braiding
under PBC with NHSE under OBC is the breaking of conventional standing-wave
conditions due to non-Hermiticity, equivalently reflected by nonzero braiding degree and persistent currents~\cite{Zhang2020}.

\subsection{Non-Hermitian skin effect and generalized Brillouin zone}
A common feature of non-Hermitian lattice Hamiltonian is the breakdown of conventional Bloch theory under
OBC, the failure of the bulk-boundary correspondence in Hermitian scenario~\cite{Yao2018,Yokomizo2019}.
In many such systems, a majority of eigenmodes localize exponentially at one boundary, namely, NHSE. Its onset can be traced back to  the dramatic spectral modifications when passing from PBC
to OBC as shown in FIG.~\ref{figphaseEnergy} (b)--(d), rendering the conventional real Bloch
wavevectors inadequate for describing the properties of the non-Hermitian bulk Hamiltonian. A modified
solution is to move to the framework of GBZ~\cite{Yao2018,Yokomizo2019}. Beyond providing the correct bulk description under OBC, the GBZ also underpins our self-healing analysis by fixing the complex momentum (and thus the decay length) of OBC bulk modes, which determines where a translating scatterer overlaps a localized profile and whether the mode lies in the gain window relevant for recovery.

\begin{figure}[htbp]
\subfigure{\includegraphics[width=\linewidth]{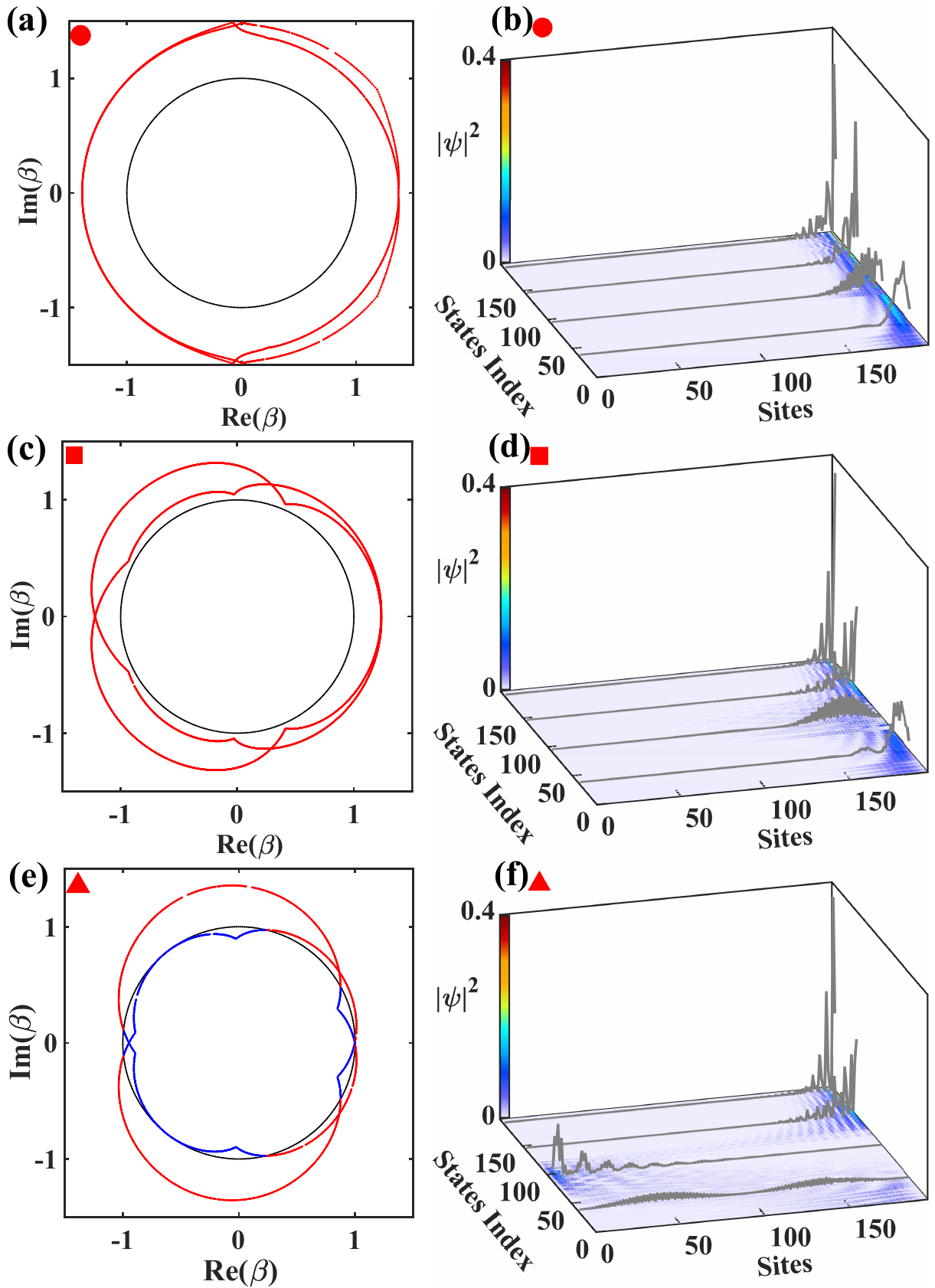}}
\caption{GBZ and distribution of eigenstates. (a), (c), (e) GBZ trajectories corresponding
    to the parameters indicated by solid circle, square and triangle indicated in FIG.~\ref{figphaseEnergy} (a),
    respectively. The points located inside the unit circle are colored blue, those located outside the unit circle
    are colored red, while the black points construct the unit circle. (b), (d), (f) Spatial distributions of the
    eigenstates under OBC at the corresponding parameters. The gray curves are used to visually represent the color
    mapping of the eigenstate-population probability.}
\label{figgbzNhse}
\end{figure}

Expanding the standard Bloch theory, $e^{ik}$ should be replaced by a complex parameter $\beta$, namely,
momentum $k$ becomes complex, thus allowing for wavefunctions $\sim \beta^{n}$ whose exponential decay
compensates for non-unitary hoppings. Concretely, we define
\begin{equation}
f(\beta,E)\;=\;\det\!\Bigl[E-H(\beta)\Bigr] \;=\;0,
\label{eqcharPoly}
\end{equation}
where $H(\beta)$ is obtained from the Bloch Hamiltonian ~(\ref{Hk}) by substituting
$e^{ik}\to \beta$~\cite{Yao2018,Yokomizo2019} in calculating the GBZ. Since our model involves up to
nearest-neighbor hopping in each sublattice sector, $f(\beta,E)$ is a polynomial of degree $4$ in $\beta$,
whose roots $\{\beta_i\}$ generically lie in the complex plane. Ordering them by magnitude,
$|\beta_1|\le |\beta_2|\le |\beta_3|\le |\beta_4|$, the GBZ is given by the closed trajectory along which
the two middle moduli become equal, $|\beta_{2}|=|\beta_{3}|$. Physically, this equal-modulus condition pairs a growing with a decaying component so the bulk amplitude neither blows up nor vanishes, thereby identifies the
dominant decay length scale of the bulk modes under OBC, as shown in FIG.~\ref{figgbzNhse} (b), (d)
and (f). See also the auxiliary GBZ formulation in Ref.~\cite{Yang2020}.

The NHSE arises because the genuine bulk eigenstates for OBC take the form
\begin{align}
\ket{\Psi_n}\,\sim\, (\beta_*)^n,
\end{align}
where $\beta_*$ lies on the GBZ rather than on the unit circle ($|\beta|=1$) in conventional Hermitian Bloch
theory. In this SU(2) SSH model, the interplay of the gauge phases $(\theta_{L}, \theta_{R})$ with asymmetric
hopping amplitudes $(t_{1},t_{2},t_{3},t_{4})$ generically warps the GBZ into an ellipse-like or more complicated
contour in the complex $\beta$-plane as shown in FIG.~\ref{figgbzNhse} (a), (c) and (e).

For the non-Abelian SU(2) SSH chain, defined in the Hamiltonian~(\ref{eqmodel}), the NHSE manifests as soon
as the hopping amplitudes and with different non-Abelian gauge angles $(\theta_L,\theta_R)$ breaks the
Hermiticity in an asymmetric manner. Concretely, our model
uses the hopping strengths $t_{1} = 0.60, t_{2} = 1.00, t_{3} = 0.80, t_{4} = 0.89$. Under the specific
parameters above, one can indeed observe in FIG.~\ref{figgbzNhse} (a), (c) and (e), that the solutions
for $\beta$ deviate significantly from the unit circle and giving
rise to strong exponential localization at the lattice boundary where non-Hermiticity is most prominent.
It can be seen in FIG.~\ref{figgbzNhse} (b), (d) and (f), this GBZ deformation induces extensive eigenstate
accumulation, predominantly localized at the boundary, which is showed vividly in movie1.

In FIG.~\ref{figgbzNhse}, panels (a), (c) and (e) show the GBZ contours for the three representative parameter
sets indicated by the circle, square and triangle in FIG.~\ref{figphaseEnergy} (a), respectively.  Points with
$|\beta|>1$ are colored red, those with $|\beta|<1$ blue, and the unit circle ($|\beta|=1$) is drawn in black.
In FIG.~\ref{figgbzNhse} (a), the GBZ lies entirely outside the unit circle, indicating a one-sided NHSE as confirmed in panel
(b) with all eigenstates localized at the right boundary.  In FIG.~\ref{figgbzNhse} (c) the GBZ approaches but remains
exterior to the unit circle, which reduces the localization strength, as evidenced by the less pronounced
boundary accumulation of the selected eigenstates in FIG.~\ref{figgbzNhse} (d).  For the third in FIG.~\ref{figgbzNhse} (e),
the GBZ crosses the unit circle, producing comparable interior and exterior segments. Correspondingly,
FIG.~\ref{figgbzNhse} (f) reveals the eigenstates localized at both ends of the chain, characteristic of a two-sided NHSE.
These results demonstrate quantitatively how varying the SU(2) gauge angle $(\theta_L,\theta_R)$ controls both the
magnitude and the direction of skin-mode localization.

Examples of one-sided NHSE are shown in FIG.~\ref{figgbzNhse} (b) and (d), where all eigenstates accumulate predominantly
at a single boundary due to asymmetric hopping. Correspondingly, in the complex-energy spectrum
(see FIG.~\ref{figphaseEnergy} (d) and the supplemental movie1), the spectral loops exhibit a twisting pattern, which is
the hallmark of two-sided NHSE, indicating that some modes localize at the left edge while others localize at the right
edge. This phenomenon arises from the interplay between the non-Hermitian hopping amplitudes $(t_1, t_2, t_3, t_4)$ and
the SU(2) gauge phases $(\theta_L,\theta_R)$.

\section{Effect of Non-Abelian Gauge Couplings on Self-Healing Dynamics}
\label{sec4}
\begin{figure}[htbp]
    \centering
    \includegraphics[width=0.9\linewidth]{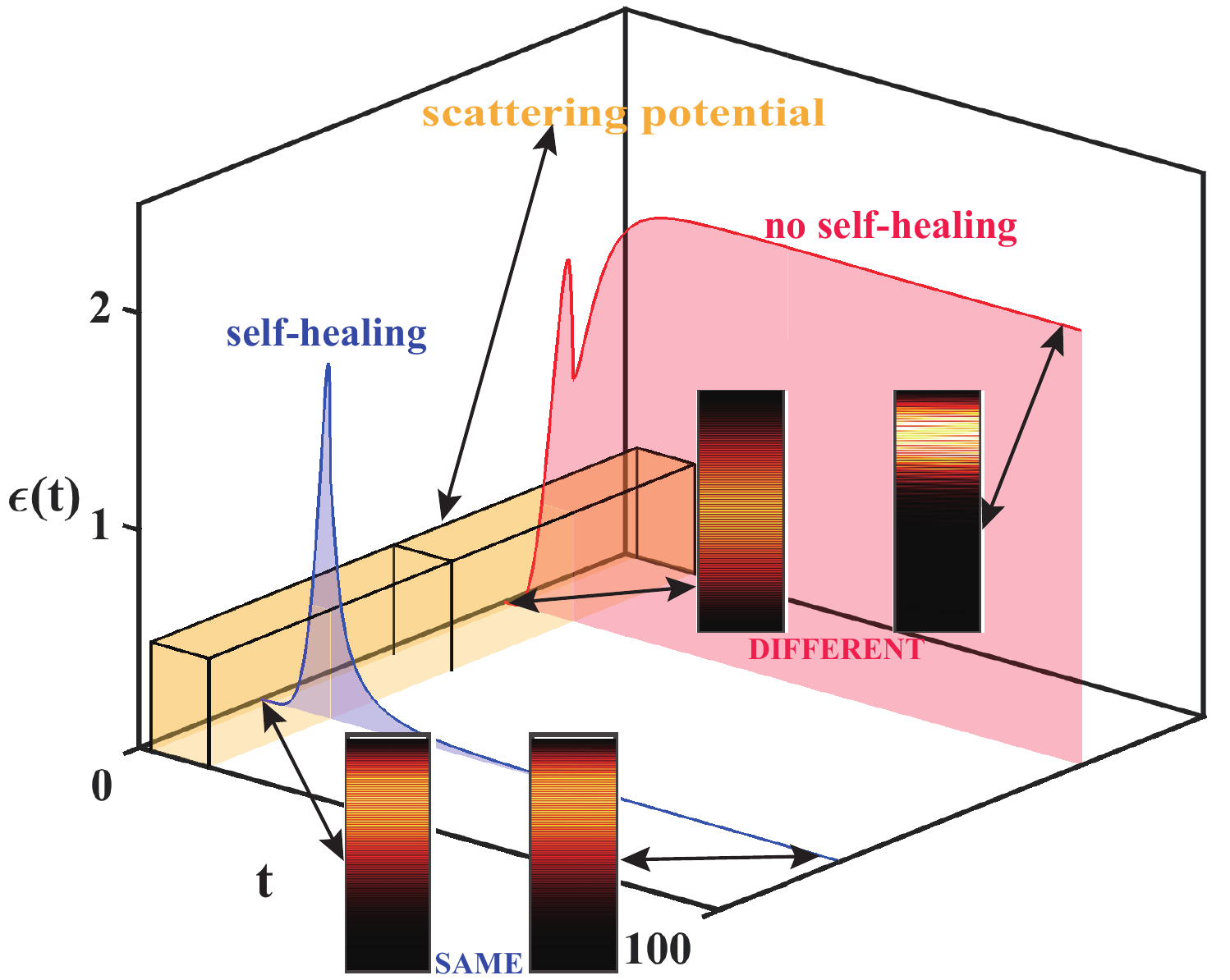}
    \caption{
    Comparison between the representative self-healing case (blue) and the no self-healing case (red) under the same scattering potential. The curves show the deviation metric $\varepsilon(t)$.
    Insets correspond to wavefunction intensity $|\psi_j|^2$ sampled at the arrow-indicated times, before and after the scattering interval. For the self-healing state, the post-scattering profile recovers and remains the same as the initial one (SAME),
    whereas the no self-healing state fails to restore its original profile (DIFFERENT).
}
    \label{fig111}
\end{figure}

In non-Hermitian lattices with topological skin modes, there exists a critical threshold
$ \text{Im}(E_{\rm th}) $ such that self-healing occurs when the imaginary part of the
eigenenergy exceeds this threshold~\cite{LonghiSelfHealinPRL,Xueeee2025}. FIG.~\ref{fig111} provides a schematic view of self-healing and no self-healing evolution. This work focuses on the
eigenstate with the largest imaginary part of its eigenenergy, $ \max \text{Im}(E) $, under OBC,
and examines whether it exceeds the critical threshold $ \text{Im}(E_{\rm th}) $. If the eigenstate corresponding
to $ \max \text{Im}(E) $ is capable of self-healing (i.e., $ \max \text{Im}(E) > \text{Im}(E_{\rm th}) $), the
system will always have at least one eigenstate capable of self-healing; otherwise, no such eigenstate exists.
This choice is natural, as the most amplified component governs the long-time dynamics and offers the clearest
diagnosis through $\varepsilon(t)$. Moreover, such high-gain modes and their self-healing behavior have already
been observed in photonic waveguide arrays~\cite{Bai2025,Ma2025}. We investigate the self-healing dynamics by supplementing the OBC Hamiltonian
$H_{\mathrm{OBC}}$ given by Eq.~(\ref{eqmodel}) with a moving, finite-width, time-windowed scattering
potential
\begin{equation}
  H(t)=H_{\mathrm{OBC}}+V(t),                                              \label{eq:Htotal}
\end{equation}
where $ V(t) $ acts diagonally on the site occupations and is switched on only during a
finite interval,
\begin{equation}
  V(t)=
  \begin{cases}
    V_0(t), & t_{\mathrm{on}} \le t \le t_{\mathrm{off}},\\[2pt]
    0,      & \text{otherwise}.
  \end{cases}                                                           \label{eq:Vpiecewise}
\end{equation}

Throughout this work, we fix the (purely imaginary) potential strength to
\begin{equation}
  V_0(t)
  = -\,i\,\Omega
    \sum_{j = j_{\mathrm{start}}(t)}^{j_{\mathrm{end}}(t)} a_j^\dagger a_j + b_j^\dagger b_j,
  \label{eq:V0}
\end{equation}
with $\Omega$ = 10. The potential always spans a block of $ N_{\mathrm{b}} = 10 $ consecutive lattice sites as
it translates on the lattice. The scattering is applied only between  $t_{\mathrm{on}}$ = 2 and $t_{\mathrm{off}}$ = 12,
namely, the total duration of the scattering potential is $ \Delta t_{\mathrm{sc}} = t_{\mathrm{off}} - t_{\mathrm{on}} = 10 $.
The system consists of $ N $ unit cells, each containing two sublattice sites $ A $ and $ B $, with each
site hosting a spin-$1/2$ degree of freedom. Hence, the full Hilbert space has dimension $ L = 4N $.
The normalized time parameter is
\begin{equation}
  \eta(t) = \frac{t - t_{\mathrm{on}}}{\Delta t_{\mathrm{sc}}},
\end{equation}
which increases uniformly from 0 to 1 during the scattering process. The leftmost site of the scattering block
then evolves as
\begin{equation}
  j_{\mathrm{start}}(t) =
  1 + \Bigl\lfloor \eta(t) \, \bigl(L - N_{\mathrm{b}} \bigr) \Bigr\rfloor ,      \label{eq:jstart}
\end{equation}
and the right edge is
\begin{equation}
  j_{\mathrm{end}}(t) = j_{\mathrm{start}}(t) + N_{\mathrm{b}} - 1.             \label{eq:jend}
\end{equation}
Eqs.~(\ref{eq:jstart}) and (\ref{eq:jend}) guarantee
$ 1 \le j_{\mathrm{start}}(t) \le j_{\mathrm{end}}(t) \le L $ for all $ t $ and realise a uniform
translation of the scattering block from the very left edge
$(j_{\mathrm{start}} = 1)$ at $ t = t_{\mathrm{on}} $ to the rightmost permissible position
$(j_{\mathrm{end}} = L)$ at $ t = t_{\mathrm{off}} $.

\begin{figure}[htbp]
\subfigure{\includegraphics[width=\linewidth]{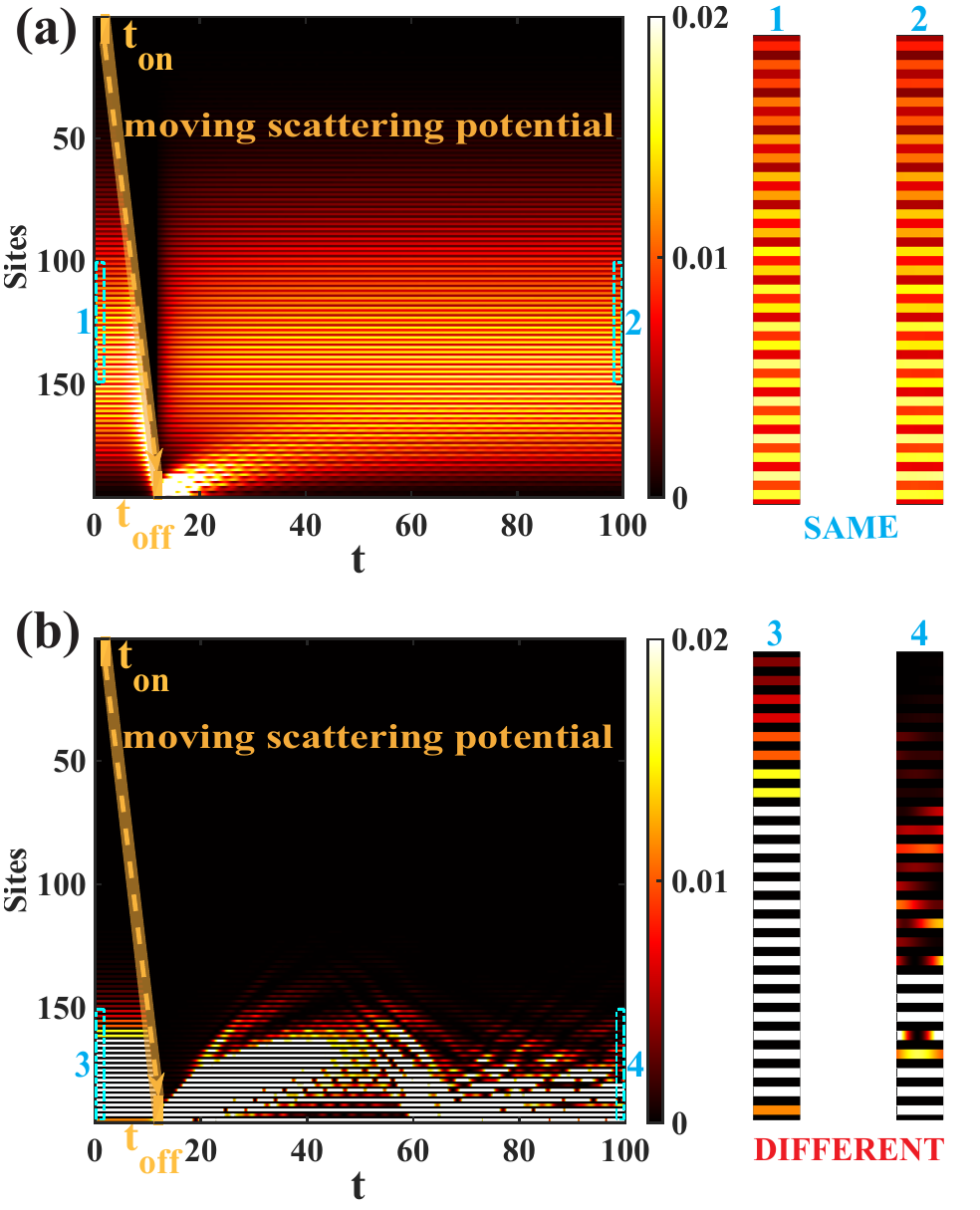}}
\caption{Wavefunction evolution from two different OBC eigenstates with the largest $\mathrm{Im}\,(E)$:
(a) with non-Abelian coupling and (b) without it.
Both cases use $t_1=0.60$, $t_2=1.00$, $t_3=0.80$, $t_4=0.89$,
with gauge phases (a) $\theta_L=1.4,\;\theta_R=0.6$ and (b) $\theta_L=\theta_R=0$.
Panels 1--4 on the right are zoom-ins of the left panels: 1 and 2 from (a) and 3 together with 4 from (b); each pair shows the unperturbed initial state and the post-scattering state. The amber dashed line marks the trajectory of the moving scattering potential.
}
\label{figselfHealing}
\end{figure}

Under the action of $V(t)$ mentioned above, we check the dynamics of self-healing with the initial wavefunction $ \Phi_E(0) $,
chosen to be the eigenstate of $ H_{\text{OBC}} $ with the maximal imaginary part $ \text{Im}(E) $ of eigenenergy. The self-healing
dynamics can be quantified by evaluating the deviation between the fully perturbed normalized wavefunction
$ \Psi(t) \sim \mathcal{T} \exp[-i \int^t d\tau H(\tau)] \Phi_E(0) $ and the ideal, unperturbed normalized state evolution
$ \Phi_E(t) \sim e^{-i H t} \Phi_E(0) $. We define the deviation as
\begin{equation}
\delta\Psi(t) = \Psi(t) - \Phi_E(t),
\label{eqDelta}
\end{equation}
and monitor its normalized overlap with the unperturbed solution via
\begin{equation}
\epsilon(t) = \frac{\|\delta\Psi(t)\|^2}{\|\Phi_E(t)\|^2}.
\label{eqdeviation}
\end{equation}

A wavefunction is said to be self-healing if $ \epsilon(t) $ vanishes as $ t \to \infty $.
In these calculations, we set $ \hbar = 1 $ and measure time in units of $ t_2^{-1} $. This
choice means that the dynamics are expressed in the units set by the inverse hopping time-scale,
facilitating comparison across different parameter sets. The results are obtained by using a fourth-order Runge-Kutta method, with explicit wavefunction normalization performed at each time step. In the simulations, we integrate over $t\in[0,100]$ with 10,000 uniform steps (i.e., $\Delta t=0.01$).

To better understand the impact of non-Abelian gauge couplings on self-healing, we first examine the
non-Abelian case (e.g., $\theta_L = 1.4, \theta_R = 0.6$).
FIG.~\ref{figselfHealing}(a) shows that after the potential is turned off at $t=12$, the wavefunction rapidly
returns to its initial spatial profile, with $\epsilon(t)$ quickly approaching zero.
By contrast, in the Abelian limit ($U_{L,R}=I,\;\theta_L=\theta_R=0$),
FIG.~\ref{figselfHealing}(b) shows a clear deviation from the original profile,
reflected in a large value of $\epsilon(t)$.

\begin{figure}[htbp]
  \centering
  \subfigure{\includegraphics[width=\linewidth]{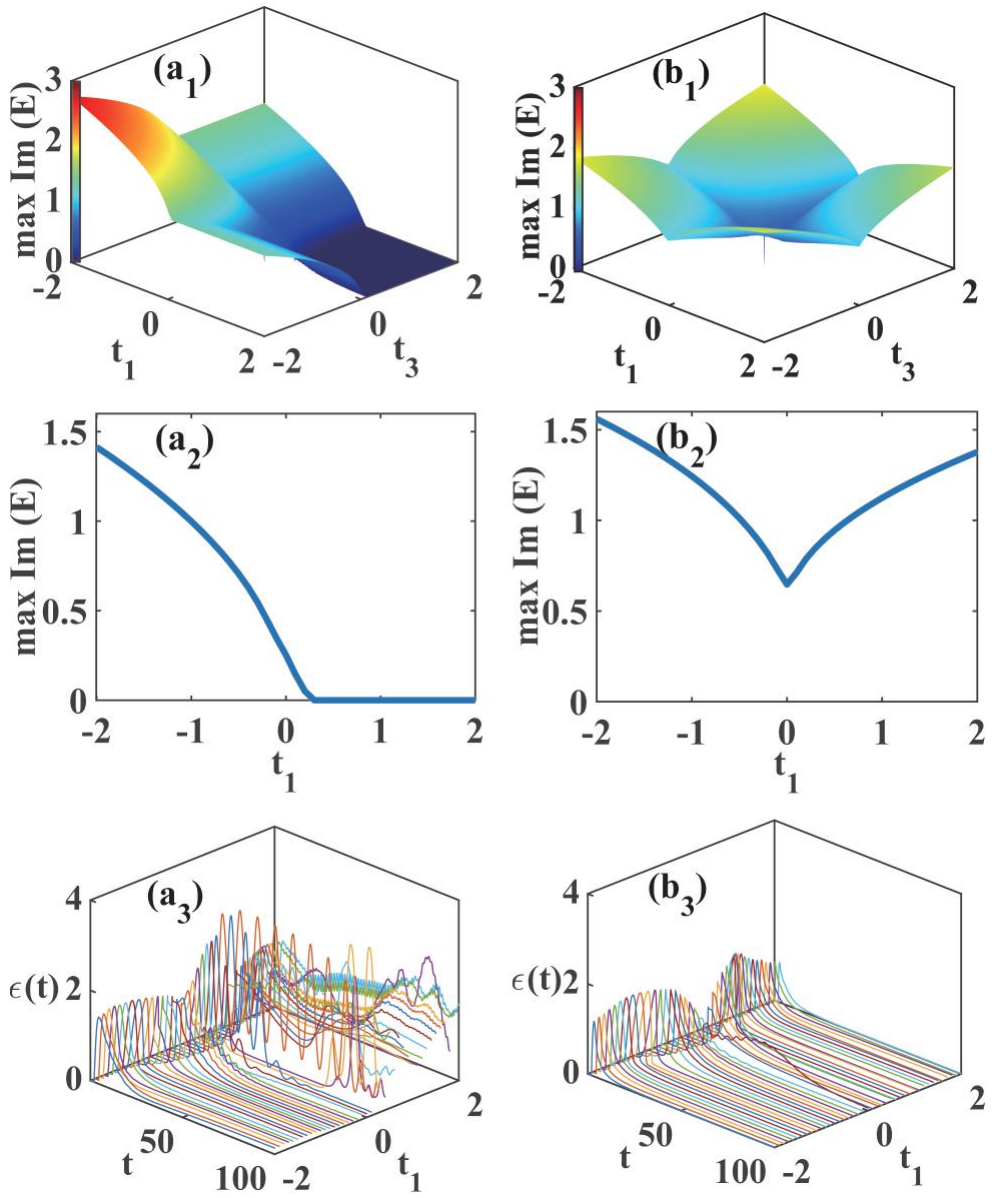}}
  \caption{($\text{a}_1$-$\text{a}_3$) correspond to the Abelian regime ($\theta_L = \theta_R = 0$), while ($\text{b}_1$-$\text{b}_3$) represent the non-Abelian regime
($\theta_L = 1.4$, $\theta_R = 0.6$), with $t_2 = 1.00$ and $t_4 = 0.89$ fixed.
($\text{a}_1$), ($\text{b}_1$): Maximum imaginary eigenvalue $\max \text{Im}(E)$ as a function of $t_1$ and $t_3$.
($\text{a}_2$), ($\text{b}_2$): $\max \text{Im}(E)$ as a function of intra-cell hopping $t_1$ for fixed $t_3 = 0.80$.
($\text{a}_3$), ($\text{b}_3$): Corresponding self-healing performance, measured by $\epsilon(t)$, as $t_1$ varies from $-2$ to $2$.}
  \label{figselfHealing11}
\end{figure}
As $ \max \text{Im}(E) $ determines the presence of self-healing, we show this imaginary part versus  $ t_1$ and
$t_3 $ in FIG.~\ref{figselfHealing11} for the Abelian and non-Abelian cases, respectively. As shown in
FIG.~\ref{figselfHealing11}($\text{a}_1$), in the Abelian limit, when $t_1$ ($t_3$) is fixed and greater than $0.3$
(a value that depends on the system size, as discussed in Appendix~\ref{appendix1}; here, $N = 49$), the largest
imaginary part of the spectrum, $\max \text{Im}(E)$, decreases monotonically as $t_3$ ($t_1$) increases and eventually
vanishes. When $t_1$ ($t_3$) is fixed and less than $0.3$, a qualitatively similar trend is observed
initially: $\max \text{Im}(E)$ decreases monotonically with increasing $t_3$ ($t_1$). However, a key difference is that
it does not reach zero, but instead saturates at a finite minimum value greater than zero. In contrast,
FIG.~\ref{figselfHealing11} ($\text{b}_1$) shows that as $ t_3 $ ($ t_1 $) increases, $ \max \text{Im}(E) $ decreases
monotonically, reaches a finite minimum greater than zero at $ t_3 (t_1) \approx 0 $, and then increases monotonically.

Next, we show the evolution of $ \max \text{Im}(E) $ when the inter-cell hopping $ t_3 $ is fixed at 0.8 in the
Abelian and non-Abelian limit in FIG.~\ref{figselfHealing11} ($\text{a}_2$) and ($\text{b}_2$), respectively.
FIG.~\ref{figselfHealing11} ($\text{a}_3$) and ($\text{b}_3$) exhibit the dynamics of $\epsilon(t)$ versus $t_1$
corresponding to FIG.~\ref{figselfHealing11} ($\text{a}_2$) and ($\text{b}_2$), respectively. When $ \max \text{Im}(E) = 0$
in FIG.~\ref{figselfHealing11} ($\text{a}_2$), the deviations $ \epsilon(t) $ in FIG.~\ref{figselfHealing11} ($\text{a}_3$)
exhibit persistent oscillations, indicating that no stable self-healing modes remain. By introducing an SU(2) gauge
field $ (\theta_L = 1.4, \theta_R = 0.6) $ qualitatively changes the behavior: $ \max \text{Im}(E) $ attains a finite minimum
at $ t_1 \approx 0 $ and then increases, as shown in FIG.~\ref{figselfHealing11} ($\text{b}_2$). FIG.~\ref{figselfHealing11} ($\text{b}_3$)
demonstrates that $ \epsilon(t) $ remains small and rapidly converges to zero over the entire interval $ -2 \le t_1 \le 2 $.

By comparing FIG.~\ref{figselfHealing11} ($\text{a}_2$) with ($\text{a}_3$), and ($\text{b}_2$) with ($\text{b}_3$), one can conclude
that for the model with $N=49$ used in this work, when $ \max \text{Im}(E) = 0 $, no eigenstates capable of self-healing are present
in the system. Conversely, when $ \max \text{Im}(E) > 0 $, the system always possesses at least one eigenstate that allows for self-healing.

In the Abelian model as shown in FIG.~\ref{figselfHealing11} ($\text{a}_1$), the contour $ \max \text{Im}(E) = 0 $ (the dark blue area),
is confined to the region where both $ t_1 \gtrsim 0.3 $ and $ t_3 \gtrsim 0.3 $, which marks the complete absence of self-healing states
in this model. In the non-Abelian model as shown in FIG.~\ref{figselfHealing11} ($\text{b}_1$), this contour touches zero at $ t_1 = t_3 = 0 $.
Everywhere else within the examined range, $ \max \text{Im}(E) $ remains positive, ensuring that at least one self-healing
mode exists. While the point $ t_1 = t_3 = 0 $ is mathematically defined, it corresponds to a trivial limit where $t_1$ and $t_3$ vanishes
entirely. As such, it falls outside the scope of physically relevant regimes considered in this work and is not further analyzed.

Even in regions where $ t_1 \lesssim 0.3 $, both Abelian and non-Abelian cases allow for the existence of self-healing states, however,
for the same value of $ t_1 $, the deviation $ \epsilon(t) $ in the non-Abelian case remains significantly smaller than
in the Abelian case (although both tend to zero), as shown in FIG.~\ref{figselfHealing11} ($\text{a}_3$) and ($\text{b}_3$). This reflects
a uniformly higher self-healing capability in the non-Abelian case.

Self-healing arises from skin-localized modes with large imaginary eigenenergies, which dominate scattering-induced excitations and restore the
spatial structure of selected eigenstates~\cite{LonghiSelfHealinPRL}. Such eigenstates with the largest imaginary parts have also been experimentally prepared~\cite{Bai2025}. We find that non-Abelian gauge couplings significantly enhance this
robustness compared to the Abelian case, enabling wavefunction recovery across a broader range of conditions. These results suggest practical
strategies for stabilizing transport in photonic or magnonic systems subject to environmental disturbances. All elements of the Hamiltonian
are experimentally accessible: asymmetric couplings have been realized in ring resonators~\cite{Liu2022LaserArrays} and magnonic
structures~\cite{Zhang2017CavityMagnonEP}; spin-dependent tunneling phases via Raman dressing~\cite{Rey2016SyntheticGauge, Guo2025DensityGauge}
and photonic birefringence~\cite{Zhan2022MetasurfaceSpin,Whittaker202115193,1Sarsen201999}; and the absorptive potential through engineered
optical loss~\cite{Pereira20246}. These capabilities suggest that the predicted effects are within reach of current or near-term experimental platforms.

\section{Conclusion}
\label{sec5}

We have introduced a non-Hermitian SSH model enriched by SU(2) non-Abelian gauge fields, leading to spin-dependent
hopping amplitudes and a rich topological structure. This framework realizes complex-energy spectral braiding
protected by chiral symmetry and supports tunable one-sided and two-sided NHSE, controllable via gauge configurations
and captured by the GBZ analysis. Crucially, we demonstrate that the presence of non-Abelian gauge fields enables
enhanced dynamical self-healing of skin-localized modes in response to time-dependent perturbations. This offers a
proposal for dynamically stabilizing edge states in non-Hermitian systems. Given the compatibility of our model
with magnonic, superconducting, atomic, and photonic platforms, the predicted phenomena, namely spectral braiding,
gauge-tunable NHSE, and self-healing dynamics, can be realized experimentally, offering new avenues for controllable
non-Hermitian topology and robust information transport. Finally, the SU(2) gauge serves as a practical control knob for robustness, enabling gauge-tunable self-healing across photonic, magnonic, cold-atom, and superconducting-circuit platforms~\cite{Bai2025}.

\section{Acknowledgements}
X. L. Zhao thanks discussions with Xingyuan Zhang, National Natural Science
Foundation of China, No.12005110, and Natural Science Foundation of Shandong
Province, China, No.ZR2020QA078, No.ZR2023MD064, ZR2022QA110.

\begin{appendix}

\section{Comparison of Self-Healing Behavior in Abelian and Non-Abelian Regimes across System Sizes}
\label{appendix1}

\begin{figure}[htbp]
  \centering
  \subfigure{\includegraphics[width=\linewidth]{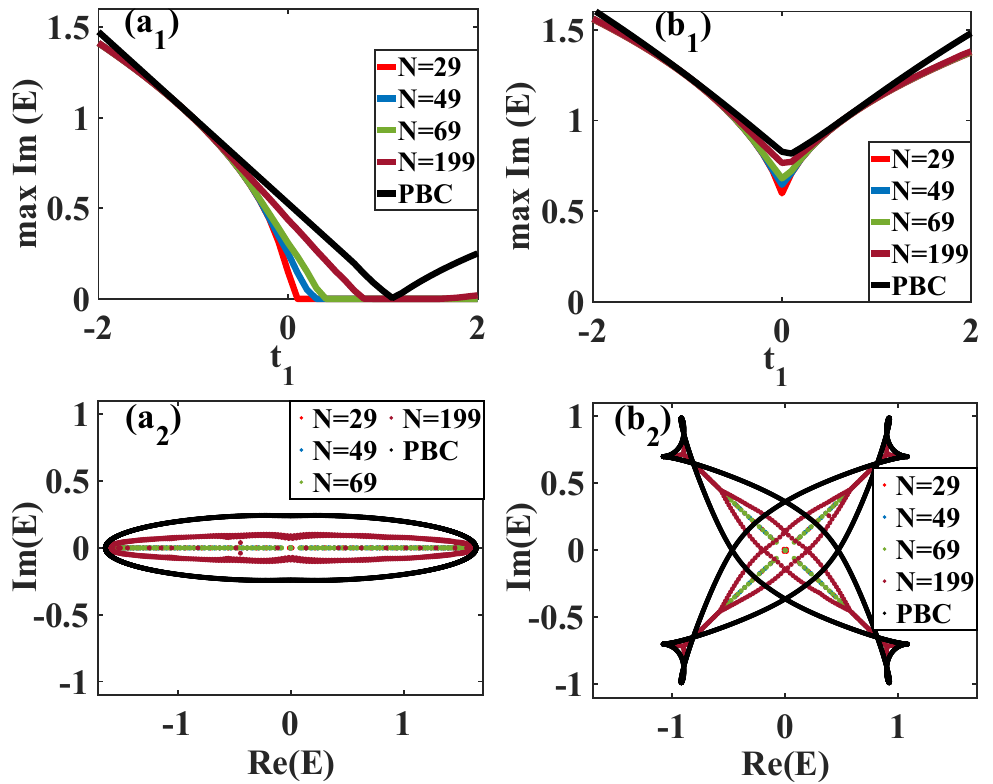}}
  \caption{($\text{a}_1$) and ($\text{b}_1$) show $\max \text{Im}(E)$ as a function of the intra-cell hopping $t_1$
  with fixed values $t_2 = 1.00$, $t_3 = 0.80$, and $t_4 = 0.89$, corresponding to the Abelian ($\theta_L = \theta_R = 0$)
  and non-Abelian ($\theta_L = 1.4$, $\theta_R = 0.6$) regimes, respectively. ($\text{a}_2$) and ($\text{b}_2$) display
  the corresponding real-space energy spectra for varying system sizes $N$, with $t_1$ fixed at $0.60$.}
  \label{s1}
\end{figure}

It should be emphasized that the turning point of $\max \operatorname{Im}(E)$, which marks a transition from decreasing to
remaining constant in the Abelian case (FIG.~\ref{figselfHealing11} ($\text{a}_1$), ($\text{a}_2$)) and from decreasing to
increasing in the non-Abelian case (FIG.~\ref{figselfHealing11} ($\text{b}_1$), ($\text{b}_2$)), shifts in position depending
on the system size $N$. Furthermore, as shown in FIG.~\ref{s1} ($\text{a}_1$) and ($\text{b}_1$), the turning point shifts to
larger values of $t_1$ (or $t_3$) as $N$ increases, indicating that the turning point is not fixed, but rather depends on the
system size.Although the condition $\max \text{Im}(E) = 0$ may still occur for increasing $N$, it does not necessarily imply
the absence of self-healing. A proper determination of self-healing should instead rely on the relative magnitude between
$ \max \text{Im}(E) $ and $\text{Im}(E_{\rm th})$.

The dependence of the turning point position on system size reflects a fundamental property of non-Hermitian systems:
unlike Hermitian systems, where the spectra under OBC and PBC are essentially identical even for finite sizes, non-Hermitian systems
exhibit strong boundary sensitivity, and the spectra under the two boundary conditions only become similar in the thermodynamic limit.
As shown in FIG.~\ref{s1}($\text{a}_2$), ($\text{b}_2$) and the movie2, as $N$ increases, the OBC spectrum gradually approaches the
PBC spectrum, and in the limit $N \to \infty$, the two spectra become nearly identical.
\begin{figure}[htbp]
  \centering
  \subfigure{\includegraphics[width=\linewidth]{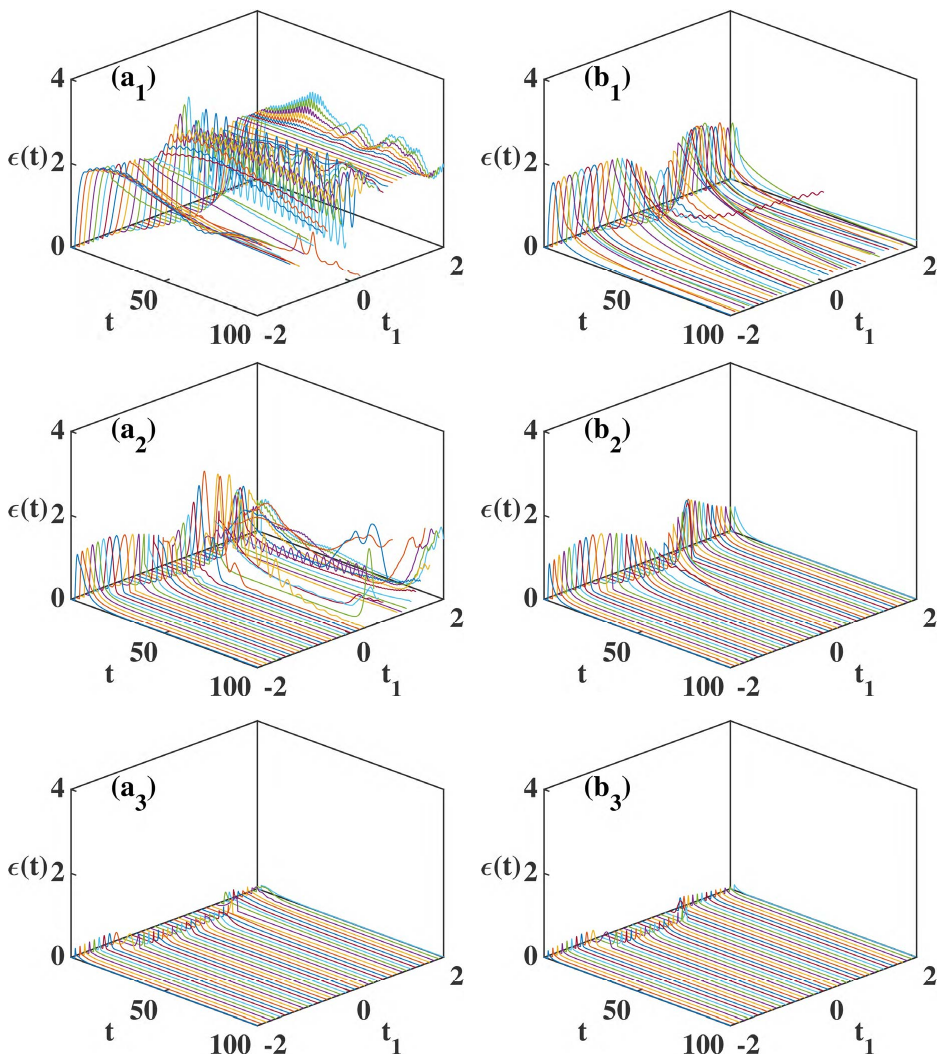}}
  \caption{Evolution of the eigenstate corresponding to the eigenenergy with the largest imaginary part as a function of
  $t_1$ (ranging from $-2$ to $2$) for different system sizes, shown for both the Abelian (panels ($\text{a}_1$)-($\text{a}_3$),
  with $\theta_L = \theta_R = 0$) and non-Abelian (panels ($\text{b}_1$)-($\text{b}_3$), with $\theta_L = 1.4$, $\theta_R = 0.6$)
  regimes. The parameters are fixed at $t_2 = 1.00$, $t_3 = 0.80$, and $t_4 = 0.89$. ($\text{a}_1$), ($\text{b}_1$): $N = 29$;
  ($\text{a}_2$), ($\text{b}_2$): $N = 69$; ($\text{a}_3$), ($\text{b}_3$): $N = 199$.}
  \label{s2}
\end{figure}

Despite these size effects, the enhancement of self-healing due to non-Abelian gauge fields remains robust. As shown in FIG.~\ref{s2},
for $N = 29$, the Abelian case (FIG.~\ref{s2} ($\text{a}_1$)) exhibits an absence of self-healing states under OBC across almost the
entire parameter range. In contrast, the non-Abelian case (FIG.~\ref{s2} ($\text{b}_1$)) retains robust self-healing behavior in nearly
all simulated configurations. For $N = 69$ (FIG.~\ref{s2} ($\text{a}_2$) and ($\text{b}_2$)), both regimes exhibit trends similar to
those in FIG.~\ref{figselfHealing11} of the main text, with the non-Abelian system consistently supporting a higher degree of self-healing.
When the system size increases to $N = 199$ (FIG.~\ref{s2} ($\text{a}_3$) and ($\text{b}_3$)), both Abelian and non-Abelian configurations
support self-healing across all parameter choices. This occurs primarily because the system length significantly exceeds the range of
the scattering potential, thereby diminishing its relative impact. Importantly, even in this regime, a comparison of the self-healing
metric at the moment the potential is applied, especially near $t_1 \approx 2$, reveals that the non-Abelian systems remain less affected
by the perturbation than their Abelian counterparts. These comparisons clearly demonstrate that the self-healing performance in the
non-Abelian regime consistently surpasses that of the Abelian case, irrespective of system size. Moreover, from
FIG.~\ref{s2} ($\text{a}_1$)-($\text{a}_3$) and ($\text{b}_1$)-($\text{b}_3$), it can be seen that for a fixed scattering potential,
its influence on the system becomes progressively weaker with increasing $N$.

\end{appendix}

\section*{Data Availability}
The \texttt{.fig}, and \texttt{.mat} files supporting the findings of this study, including all numerical results shown in FIGs.~\ref{figphaseEnergy}, \ref{figgbzNhse} and~\ref{figselfHealing}--\ref{s2}, as well as the codes for constructing the GBZ and for time-evolution simulations, are openly available at Zenodo: \href{https://doi.org/10.5281/zenodo.17052695}{10.5281/zenodo.17052695}~\cite{Miao2025data}.

\end{document}